\documentclass[final,5p,times,twocolumn]{elsarticle}

\usepackage{amsmath,amssymb,stmaryrd}
\usepackage{amsthm}
\usepackage{array}
\usepackage{siunitx}

\begin{document}
\newcommand{\CPMD}{\textsc{cpmd}}
\newcommand{\citeme}{$^{\text{XXX}}$}
\begin{frontmatter}



\title{
Integrating State of the Art Compute, Communication, and Autotuning\\
Strategies to Multiply the Performance of the Application Programm CPMD\\
for \textit{Ab Initio} Molecular Dynamics Simulations
}

\author[1,2]{Tobias Kl\"offel}
\ead{tobias.kloeffel@fau.de}
\address[1]{Interdisciplinary Center for Molecular Materials (ICMM) and
  Computer-Chemistry-Center (CCC),\\ Friedrich-Alexander-Universit\"at
  Erlangen-N\"urnberg (FAU), N\"agelsbachstra{\ss}e 25, 91052 Erlangen,
  Germany}
\address[2]{High Performance Computing Group, Regional Computing Center
  Erlangen (RRZE),\\ Friedrich-Alexander-Universit\"at Erlangen-N\"urnberg
  (FAU), Martensstra{\ss}e~1, 90158~Erlangen, Germany}

\author[3]{Gerald Mathias}
\address[3]{Leibniz Supercomputing Centre of the Bavarian Academy of Sciences
            and Humanities, Boltzmannstr.~1, 85748 Garching, Germany}
\ead{gerald.mathias@lrz.de}

\author[1]{Bernd Meyer\corref{cor1}}
\cortext[cor1]{Corresponding author.}
\ead{bernd.meyer@fau.de}

\begin{abstract}
We present our recent code modernizations of the of the {\em ab initio} molecular dynamics
program CPMD (www.cpmd.org) with a special focus on the ultra-soft pseudopotential
(USPP) code path.
Following the internal instrumentation
of CPMD, all time critical routines have been revised
to maximize the computational throughput and to minimize the communication overhead
for optimal performance.
Throughout the program
missing hybrid MPI+OpenMP parallelization has been added to optimize scaling. 
For communication intensive routines, as the multiple distributed 3-$d$ FFTs of
the electronic states and  distributed matrix-matrix
multiplications related to the $\beta$-projectors of the pseudopotentials,
this MPI+OpenMP parallelization now overlaps computation and communication.
The necessary partitioning of the workload is optimized by an auto-tuning algorithm. 
In addition, the largest global {\tt MPI\_Allreduce} operation has been replaced by highly
tuned node-local parallelized operations using MPI shared-memory windows 
to avoid inter-node communication.
A batched algorithm for the multiple 3-$d$ FFTs improves the throughput of 
the {\tt MPI\_Alltoall} communication and, thus, the scalability of the
implementation, both for USPP and for the frequently used
norm-conserving pseudopotential code path.
The enhanced  performance and scalability is demonstrated on a
mid-sized benchmark system of 256 water molecules and further water systems
of from 32 up to 2048 molecules.
\end{abstract}


\end{frontmatter}

\section{Introduction}

The field of {\em ab initio} molecular dynamics (AIMD) has been enormously  growing 
over more than two decades providing exciting new insights to chemistry and
materials science at atomic resolution.
In AIMD atoms are propagated in time according to 
Newton's equations of motion to study physical properties and chemical reactions 
of molecules and condensed phase systems at finite temperature.\cite{Hutter05,Marx2009} 
The necessary energies and forces are derived from first principles 
quantum chemistry, whose enormous computational effort has made it 
a key application in high-performance computing (HPC).

One of the most successful implementations of AIMD is the CPMD\cite{CPMD43}
code, which has been available on almost all major HPC platforms since the
late 1990'. The code is MPI\cite{MPI31} parallel and offers  a 
OpenMP\cite{OpenMP45} parallelization on top for many time critical routines.

A key factor to its success always was the porting and optimization 
for new compute architectures, like vector machines or the multi-core architecture
of the IBM Blue Gene series.
More recent porting efforts focus on the support of accelerators, namely GPUs \cite{Weber2016},
which are one aspect of a general trend in  high performance computing.
In the last decade, the compute power but also the complexity of 
the nodes has grown much stronger than the speed of the interconnects.
On current supercomputers each node typically hosts dozens of of cores with
strong vector units, and the cores may be complemented by GPU accelerators.
The intra-node communication between these units is by an order of magnitude 
faster than the inter-node network.

The underlying method of CPMD is
Kohn-Sham density functional theory (DFT), which describes the electrons
through singly or doubly occupied electronic states. 
In CPMD these states are
expanded in a plane-wave (PW) basis and propagated via the Car-Parrinello extended
Langrangean technique. For details of the method and the implementations see
\cite{Hutter05,Marx2009}.

Some properties of the electrons are calculated efficiently in momentum space,
other operators are diagonal in real space. CPMD uses 3-$d$ fast Fourier
transforms (FFT) to switch back and forth between these two representations
for a most efficient computation \cite{Hutter05,ANDREONI2000819,HUTTER20051}.
Thus, the key objects of the method are the grids in real and momentum space
that are used to represent the electron density in the simulation cell, the
core potentials and the electronic states.

CPMD uses so-call pseudopotentials (PP) to describe the chemically inert core
electrons and to limit the calculation to the valence electrons \cite{HellmannPP}. 
This restriction  requires a much smaller plane-wave basis than one would need for
an all-electron calculation. For the frequently used norm-conserving PP (NCPP)
this results in real space grids of a few hundred points per dimension for the
FFT. The resolution of the grid can be further reduced by at least 40\,\% by
using Vanderbuild ultra-soft PP (USPP) \cite{Vanderbilt90}.

The  USPP method  is formulated in $N$ electronic states $\big\{|\phi_i\rangle\big\}$,
which fulfill the constraint
\begin{equation}
\label{eq:GON}
<\phi_i|S|\phi_j>=\delta_{ij}
\end{equation}
by means of the  non-local overlap operator
\begin{equation}
\label{eq:S}
S = 1+\sum\limits_{I,mn} q^I_{mn} |\beta^I_m \rangle \langle\beta^I_n |,
\end{equation}
which depends on the $\beta$-projectors $\big\{|\beta^I_m\rangle\big\}$ of the atoms $\{I\}$ and  the integrals
\begin{equation}
q^I_{mn}=\int\mathrm{d}\mathbf{r} Q^I_{mn}(\mathbf{r})
\end{equation} 
of the augmentation functions $Q^I_{mn}(\mathbf{r})$ provided with the USPPs.\cite{Vanderbilt90,Laasonen93}. 
These augmentation functions are also required to compute the total electron density
\begin{equation}
  \label{eq:density}
    n(\mathbf{r})=\sum\limits_{i}\left[|\phi_i(\mathbf{r})|^2+ \sum\limits_{mn,I}Q^I_{mn}(\mathbf{r})\langle\phi_i|\beta^I_n\rangle\langle\beta^I_m|\phi_i\rangle\right],
\end{equation}
which  contains contribution from the projections of the $|\phi_i\rangle$ onto the USPP.

The non-local overlap operator \eqref{eq:S} leads to the generalized eigenvalue problem
\begin{equation}
  \label{eq:hpsi}
  H|\phi_i\rangle = \epsilon_i S|\phi_i\rangle
\end{equation}
for the time independent Schr\"odinger equation with the Hamiltonian
\begin{equation}
  \label{eq:hamiltonian}
  H = -\nabla^2+V^\mathrm{eff} +\sum\limits_{mn,I}D^I_{mn}|\beta^I_{m}\rangle\langle\beta^I_n|,
\end{equation}
where the last term containing
\begin{equation}
  \label{eq:newd}
  D^I_{mn} = D^{(0)}_{mn} +   \int \,\mathrm{d}\mathbf{r}V^\mathrm{eff}(\mathbf{r})Q^I_{mn}(\mathbf{r})
\end{equation}
 describes the non-local part of the potential with given parameters
 $D^{(0)}_{mn}$.

Because the generalized constraint \eqref{eq:GON} is difficult to handle in
the  extended Lagrangean dynamics of  CPMD \cite{Laasonen93}  a set of
orthogonal orbitals $\big\{|\psi_i\rangle\big\}$ is used,\cite{Hutter95} which
yield the original
\begin{equation}
\label{eq:Tmat}
|\phi_i\rangle=\sum_j|\psi_j\rangle T_{ij},
\end{equation}
by the inverse root $T=O^{-1/2}$
of the overlap
\begin{equation}
  \label{eq:csmat}
  O_{ij}=\langle\psi_i|S(\{\mathbf{R}_I\})|\psi_j\rangle.
\end{equation}
Potential energies and forces are still calculated with respect to the
$\big\{|\phi_i\rangle\big\}$, which requires frequent transformations between
these two representations.

In CPMD the orthonormality of the $\big\{|\psi_i\rangle\big\}$ is not enforced
by the full set of $N^2$ coupled constraints  $\langle\psi_i|\psi_j\rangle-
\delta_{ij}=0$, but just by the $N$ diagonal constraints
$\langle\psi_i|\psi_i\rangle- 1=0$, which are decoupled and, thus, their
Langrangean multipliers are easy to determine. On the downside orthogonality
has to be enforced from time to time by solving an  $N^2$  eigenvalue
problem.\cite{Hutter95}

Using USPP, each atomic PP requires about twice as many $\beta$-projectors  as
NCPP to map the interaction of the atomic core with the electronic states
\cite{Laasonen93}. The larger number of $\beta$-projectors 
together with the additional terms appearing in the
equations above, the USPP scheme requires some extra computation.
Ideally, this should be more than compensated by the lower grid resolution and 
much smaller plane wave basis set (roughly a factor of eight).
Also from a parallelization standpoint, bearing in mind the above discussion 
on the fast growth of the nodes computational power,
the USPP approach seems favorable compared to NCPP, since it reduces the amount of communication at the 
expense of some extra computation.
However, in the current implementation of CPMD the USPP code branch
is only slightly faster than NCPP and displays rather poor scaling.
Nevertheless, USPP calculations in CPMD have been indispensable for systems with a
large number of electrons because the much smaller basis set
largely reduces memory requirements \cite{Laasonen93}.

The heart of data parallelization in CPMD is a 1-$d$ domain decomposition of the
FFT grid, i.e.\ the $yz$-planes of the electron density are distributed to 
the MPI tasks (see \cite{Hutter05,Marx2009} for details).
Thus, the few hundred grid points per dimension is an effective limit 
for the number of MPI tasks that can be efficiently used in a calculation.
A decade ago, this was not critical on most HPC
platforms, but with the emergence of powerful multi- and many-core compute nodes
pure MPI parallelization limits the achievable performance of the code. 
CPMD already can be operated in a hybrid MPI+X mode, where additional 
thread parallelization is introduced via performance libraries 
(BLAS, LAPACK, FFT)\cite{Blackford02,Anderson90,Press88} and explicit 
OpenMP\cite{OpenMP50} directives. However, not all code paths are 
equally well thread parallelized and the efficiency of CPMD in hybrid mode 
often falls behind that of pure MPI.
This problem is particularly pressing for simulations that use USPP 
due to the much smaller FFT grid.
Concomitantly, USPP use a much smaller spacial grid and reach 
the scaling limit in a pure MPI mode  earlier.
On top, the hybrid parallelization scheme is not as well maintained for 
the USPP code path as for norm-conserving PPs.

Recently, a second level of MPI parallelization was introduced in CPMD to enhance
the scalability \cite{Weber2014}. In the so-called {\tt cp\_groups} parallelization, 
the MPI tasks with their associated memory shares are replicated $g$ times. 
Each of these $g$ {\tt cp\_groups} computes the same workload per default. 
For speed-up, the members with the same index in different   {\tt cp\_groups}  
do work-sharing and data synchronization among each other. 
Originally, this was implemented to distribute state pairs for exact exchange 
and to parallelize the FFT routines over the electronic states. 
For all USPP specific routines these  {\tt cp\_groups} parallelization 
has been missing completely.

In this paper we discuss our latest revision of the CPMD code to  alleviate
the parallelization and performance bottlenecks of the USPP code path.
It builds upon our intermediate results presented at the Supercomputing Conference 2018.\cite{posterSC18}
On node level our effort aims to fully exploit the powerful compute units
of current machines and make use of the fast intra-node communication with hybrid 
MPI+OpenMP parallelization strategies.
Furthermore, we address the time critical parallel
tasks, namely the distributed matrix-matrix multiplication to calculate the overlap
of the electronic states with the $\beta$-projectors and the 3-$d$ FFT of the
electronic states. 
For both routines, respectively, we introduce overlapping computation and communication
to reduce the communication overhead. The required data partitioning strategy is determined by and 
auto-tuning algorithm. 
Where possible, inter-node communication is avoided by node local parallelization, using e.g.
 MPI shared-memory windows.
To achive maximum scaling  we complemented compute routines 
by a {\tt cp\_groups} parallelization scheme where applicable.

The following Methods section gives an overview of the systems used for benchmarking before
we describe details of the code changes in the section Optimization.
The benefits of our efforts are discussed in the section Results and a short Summary and Outlook concludes the paper.

\section{Methods}

Starting point for our optimization was the development version of CPMD 
(current release version is 4.3). CPMD brings its own instrumentation to measure
routine timings inclusive and exclusive the timing of subroutines.  For the
optimization of selected routines we have used this instrumentation to measure 
the timing of individual code blocks, as well.
Furthermore, CPMD measures the amount of data and the bandwidth 
in MPI communications, which supported the optimization of communication.
The external tools used for performance measurements and optimization were 
likwid\cite{Treibig2010}, PerSyst\cite{Guillen2014} and the {\tt optreport}
option of the Intel FORTRAN compiler,\cite{ifort2019} which helped us to check
if certain loops have been SIMD vectorized.

We will exemplify our optimization progress on a benchmark system of 256
water molecules, which lies in a medium system size range between 500 and 2000
electronic states used in many state of the art  AIMD publications
\cite{Heyden2010,Hassanali2014AqueousSS,Gaiduk2018,Rozsa6952,Zheng}. 

Molecules were simulated enclosed in a $19.734^3$\,\AA$^3$ periodic box with a
PW cut off of 25\,Ry and a time step of 6\,a.u., employing the PBXE functional\cite{PBE96}
including Grimme's D3 correction\cite{Grimme2006} for long range dispersion interactions. 
This system comprises 1024 electronic states, which are expanded in 54564 PW
basis functions. The USPP  of the Oxygen atoms contain eight projectors each;
Hydrogen atoms covers one $\beta$-projector, which leads to a total of 2560  $\beta$-projectors.
Hydrogen masses were set to 2 a.m.u.\ and the fictitious electron mass was 700 a.u.
The CPMD code was compiled with the Intel compiler suite 2019, including Intel
MPI and Intel MKL \cite{ifort2019,MKL2019,impi2019}. The latter was used both
for BLAS/LAPACK calls and for FFT. ELPA 2019.05 was compiled with the same
Intel compilers\cite{Marek2014}.

For comparison with the NCPP code path, we also benchmarked the same 256 water molecule system with
the NCPP implementation of CPMD. Here we used Trouiller-Martins
norm-conserving PPs \cite{Troullier91} with a PW cutoff of 80\,Ry, which
results in 313126 PW basis functions. Since only the 2S-spin channel of Oxygen needs
a $\beta$-projector, 256 $\beta$-projectors are included in the NCPP calculation.

We also conducted scaling tests with system sizes 32, 64, 128, 256, 512, 1024
and 2048 water molecules, respectively.
The initial simulation cells are taken from CP2K.org \cite{cp2k.org}.

For the performance measurements we have run 1000 steps of Car-Parrinello MD
and evaluated the average time per step. IO was excluded because it is
typically not a bottleneck for production runs but reading/writing the restart
(checkpoint) files can significantly add to the timings of the benchmark
runs. We checked the reproducibility of measured timings by repeating the
benchmark runs and found standard deviations of less than 0.5\,\%, which was
sufficient for our purposes.
Newly implemented auto-tuning algorithms were performed during the 
initialization phase and do not contribute to the overall timing.

The benchmarks simulations were run on SuperMUC-NG, whose nodes feature 2
$\times$ 24 cores Intel\textsuperscript{\textregistered} Skylake Xeon Platinum
8174 processors running at 2.3\,GHz and 96 GB of memory (80 GB usable). The
nodes of an island are interconnected by a fully non-blocking fat tree
100~Gbit/s Intel\textsuperscript{\textregistered} OmniPath network. 
For our benchmarks we always constrained the
nodes to be on the same island. 
Up to 384 nodes were used for a single simulation.
For hybrid MPI/OpenMP runs we benchmarked  1, 2, 3, 4, 6, 8, 12, or 24 OpenMP
tasks per MPI thread to evenly fill each node. {\tt cp\_groups}
parallelization was benchmarked with up to eight {\tt cp\_groups}.

For comparison, we benchmarked both the original CPMD 4.3 code and our revised
code. For the former, however, we omitted the
 {\tt cp\_groups} parallelization in the USPP benchmarks, 
since it was only implemented for the FFT.

\section{Optimization}
\label{sec:optimization}

Working through the list of the most time-consuming routines in CPMD we
managed to eliminate many bottlenecks.
A first focus lay on the node-level optimization for enhanced performance.
As we show below, some routines were speed up by one or two orders of
magnitude through performance libraries,
OpenMP parallelization and optimization of memory access patters, such that
these routines became less important or even negligible in the overall
timing.

The second focus was to improve the scalability within the target hybrid MPI+OpenMP parallelization.
Here, for the two most time-critical routines we have revised the respective parallel communication pattern.

\subsection{Overlap of $\beta$-projectors}
\label{ssec:ovl}
The largest computational effort with respect to matrix-matrix multiplication
and a major computational bottleneck in the USPP code path of CPMD is the
computation of the overlaps

\begin{equation}
\label{eq:fnl}
F^{\mathrm{NL}}_{I,m,i} = \langle \beta^I_m| \phi_i\rangle\quad \mathrm{and} \quad 
\tilde F^{\mathrm{NL}}_{I,m,i} = \langle \beta^I_m| \psi_i\rangle
\end{equation}
of the $\beta$-projectors and  the two sets of  electronic states,
which are  core quantites for the USPP code path due to their size.
It requires a summation over the PW coefficients, who, for the electronic
states, are distributed among the MPI tasks. Thus, the partial sums to
$F^{\mathrm{NL}}$ from each task are combined by an {\tt MPI\_Allreduce} operation.

The computation of these overlaps is implemented in the routine {\tt rnlsm1},
which calls  {\tt DGEMM} for each $\beta$-projector of each atomic species.
 {\tt rnlsm1} is  called twice in CPMD 4.3, once to calculate the projections on the
orbitals $\big\{|\psi_i\rangle\big\}$ used for the dynamics and a second time to calculate
the projections on the orbitals  $\big\{|\phi_i\rangle\big\}$ used for evaluating ionic
forces and electronic gradients.
Since these two sets of orbitals are related by the transformation
\eqref{eq:Tmat} we have eliminated the second call to  {\tt rnlsm1} and
compute the second projection by the transform

\begin{equation}
\label{eq:fnlt}
F^{\mathrm{NL}}_{I,m,i} = \sum_j  \tilde F^{\mathrm{NL}}_{I,m,j} T_{ij}.
\end{equation}
For this new routine {\tt rottr\_fnl} we have chosen a node-local
parallelization, where the work is distributed among the MPI tasks of each
node, since the overall computational effort is small but not negligible.
The node local tasks share data in  a MPI shared-memory window, and, thus,
avoid communication.

To  optimize the remaining singular call to {\tt rnlsm1}, we pack all
$\beta$-projectors in a single matrix.
A single call to {\tt DGEMM} now is most efficient; in our benchmark system it has
the large dimensions $M=2560$, $N=1024$, $K=54564$, where the inner dimension
$K$ is the number of PW, which are distributed over the MPI tasks.
The partial contributions are summed over the tasks
by {\tt MPI\_Allreduce} {\em after} the computation, which amounts to  20\,MB 
in our benchmark and hampers scaling. To improve the scaling and the
overall timing  the new implementation  subdivides the $\beta$-projector
matrix into $n$ buffers, which serve to overlap computation and communication.

\begin{figure}
\center{\includegraphics[width=3.5cm]{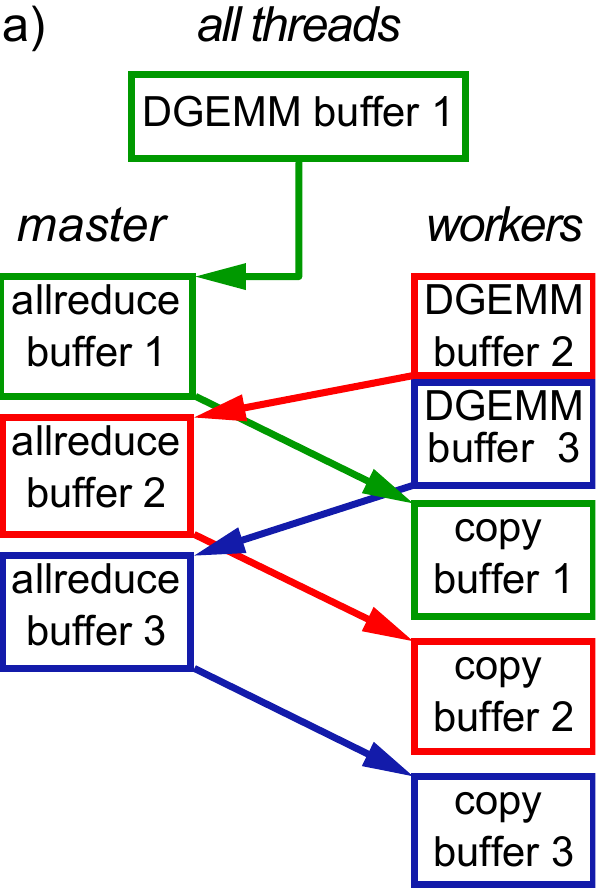}}
\caption{
\label{fig:rnlsm-flow}
Revised algorithm for {\tt rnlsm1} for $n=3$ implementing overlapping computation and communication.}
\end{figure}

Figure \ref{fig:rnlsm-flow}  sketches the resulting algorithm for the choice
$n=3$. First, all threads compute the local contribution of buffer 1 to
$F^{\mathrm{NL}}$. This is done to reduce the impact of splitting off a
complete thread and dedicate this thread to the MPI communication, which is
done after performing the first {\tt DGEMM}.
The master thread then conducts the {\tt
  MPI\_Allreduce} operation, while the remaining worker threads continue
evaluating buffers 2 to $n$. When the communication is finished, the buffer
content is copied to the internal data structure in CPMD. To enable further
scaling, we adopted the {\tt cp\_groups} parallelism to distribute the
$\beta$-projectors between the groups.

Choosing an optimal $n$ and the relative buffer sizes of buffer 1 and buffers
2 to buffer $n$ strongly depends on the number of threads, the performance of
the processor and network architecture. Therefore, we implemented an
auto-tuning algorithm, which measures the relative timing of computation and
communication  during the first few MD steps and  chooses an optimal set-up.

\begin{figure}
\includegraphics{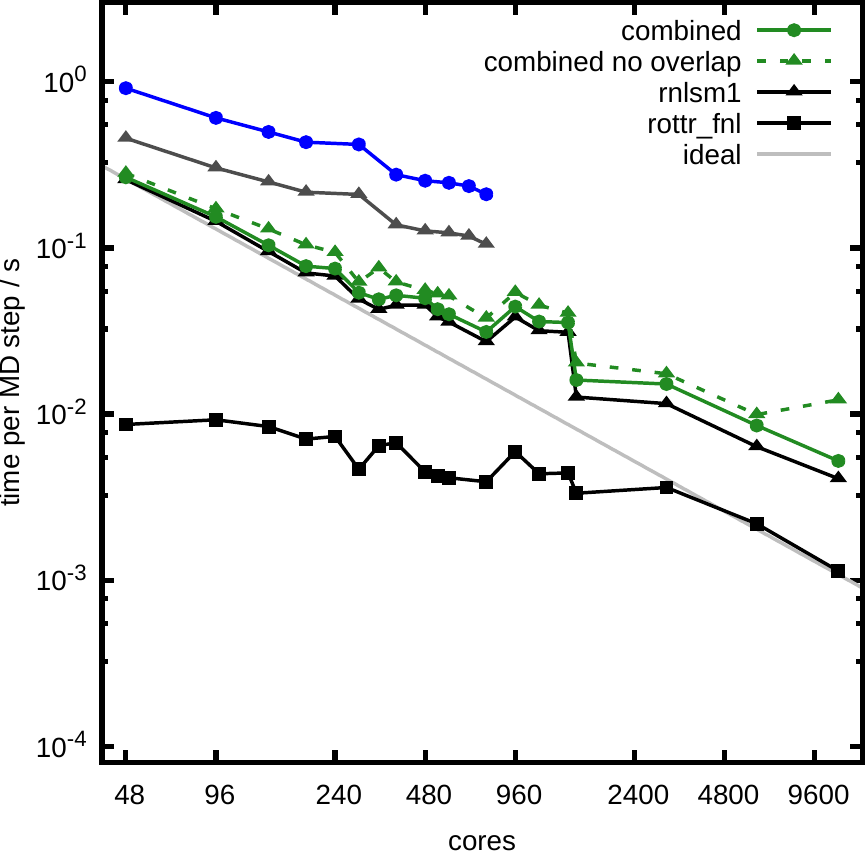}
\caption{
\label{fig:rnlsm1}
Routine specific and combined time spent to compute $\tilde F^{\mathrm{NL}}$
and $F^{\mathrm{NL}}$. The grey line represents the routine {\tt rnlsm1} of
CPMD 4.3, which iscalled twice per time step; the combined time is shown in
blue. The revised implementation calls {\tt rnlsm1} once per MD step (black
triangles) and substitutes the second call with {\tt rottr\_fnl} (black
squares). Green shows the total time of the new implementation, green dashed
shows the resulting timings if {\tt rnlsm1} is used without overlapping
communication and computation.
}
\end{figure}

Performance and scalability improvements of the revised algorithm are
presented in Figure~\ref{fig:rnlsm1}. Here, the node level optimization of 
the {\tt rnlsm1} routine yields a speedup of about 1.7 on a single node (48
cores). Up to 8 nodes (384 cores)  the new algorithms displays excellent
scaling and is more than 3.0 times faster than the original
implementation. Since the node-local rotation is almost negligible the overall
performance improvement is about a factor of 5.3. 
Beyond 8 nodes the overall timing reaches a plateau until at 32 nodes the first data
point with {\tt cp\_groups} parallelism is presented, leading to a sizable improvement.
From Fig.~2 it may look appealing to use
{\tt cp\_groups} parallelization already for fewer nodes, but
it requires additional synchronizations events throughout the code.
Even though we have minimized this synchronization of the {\tt cp\_groups},
the overall performance of the code  benefits from this additional parallelization level
only at 1536 cores and beyond for our test system, where it enables further scaling of the algorithm.

The
choice of replacing the second call to {\tt rnlsm1} with the node local {\tt
  rottr\_fnl} is clearly beneficial as the total time is now dominated by the
single {\tt rnlsm1} call. Also the parallelization of {\tt rottr\_fnl} with
an MPI shared memory window is working exceptionally well. Despite of different 
numbers of OpenMP threads at the various node counts, the routine shows 
an almost constant timing. 
Due to {\tt cp\_groups} parallelism the node local workload is reduced at core counts
beyond 3072 cores and the timing is sped up.
For a detailed comparison of number of cores and OpenMP
threads per MPI task see Tab.~\ref{tab:results} below. 
For larger systems, however, this scaling limit will manifest much later
because the computational effort grows faster with the system size than the
communication.
The overlapping computation and communication is
benefitail already upwards from two nodes and does not add any overhead for a single node.

\subsection{Ionic Gradients of $\beta$-Projectors}
\label{ssec:Ions}
In contrast to Born-Oppenheimer MD, where ionic forces are evaluated once
after an self-consistent field (SCF) cycle,  Car-Parrinello MD  requires
ionic forces together with each update of the wave function. Therefore, force
evaluation is  a dominant part of the overall computational time.  For the
forces due to the $\beta$-projectors it requires the quantity

\begin{equation}
\label{eq:dfnlt}
\mathrm{ d}F^{\mathrm{NL}}_{I,m,i} = \langle \nabla_I\beta^I_m|\phi_i\rangle,
\end{equation}

which is calculated in subroutine {\tt rnlsm2}. 

Finally, subroutine {\tt rnlfl}  calculates its contribution to the ionic forces. 
In the original implementation of
   {\tt rnlsm2} each MPI task holds a slice of
array $\mathrm{ d}F^{\mathrm{NL}}_{I,m,i}$  so that  {\tt rnlfl}  evaluates
forces due to these slices in parallel. Still, the effort is sizeable as  {\tt
  rnlfl} uses an eight-fold nested loop without any OpenMP
parallelization. Thus, both  {\tt rnlsm2}  and routines  {\tt rnlfl} are
equally time consuming in CPMD 4.3, where the scaling of  {\tt rnlfl}  was
slighly better.

In a first step,\cite{posterSC18} we have applied the same optimizations to {\tt rnlsm2} 
as to {\tt rnlsm1}
described in Section \ref{ssec:ovl}.
This lead to a similar speed up and scaling, but still 
the cost for  {\tt rnlfl} was sizeable.
Close inspection of {\tt rnlfl}  revealed that a large portion of the computational effort
can be expressed by a matrix-matrix multiplication of $F^{\mathrm{NL}}$  
to the Hamilton matrix
\begin{equation}
\label{eq:hmat}
H_{i,j} = \langle \phi_i|H| \phi_j\rangle.
\end{equation}
This product is not only required in {\tt rnlfl} but also in the routine {\tt nlforce},
which computes the gradients of the electronic states and will be discussed 
in  Section \ref{ssec:nlforce} below.
We have implemented this matrix-matrix multiplication in the new routine 
{\tt rotate\_fnl},
which uses the same parallelization strategy as in {\tt rottr\_fnl}.
Using this intermediate quantity the  revised {\tt rnlfl} now is OpenMP parallel, 
uses SIMD vectorization, and is at node level about two orders of magnitude faster. 
Because of this massive speed up of {\tt rnlfl} its parallelization across MPI tasks 
is no longer necessary and we discard it. 
As a result, the reduction operation of $\mathrm{ d}F^{\mathrm{NL}}$ across 
the MPI tasks can be skipped, which makes {\tt rnlsm2} communication free
in the USPP code path. 
For our benchmark system this discards an {\tt MPI\_Allreduce} operation 
of $3\times 20$ MB in size.

\begin{figure}
\includegraphics{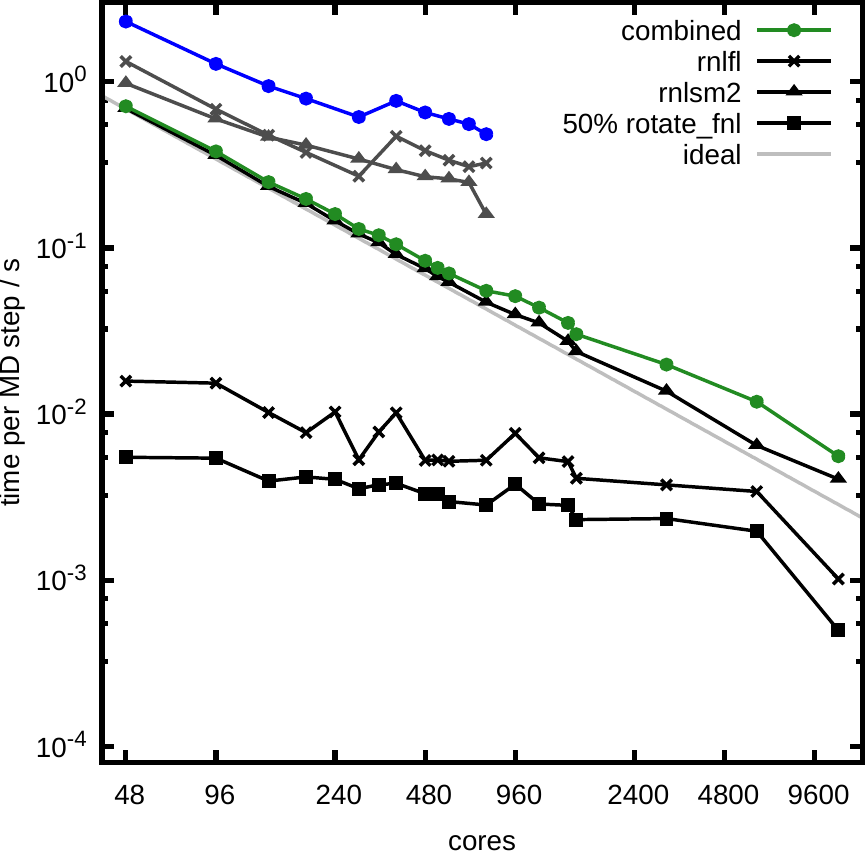}
\caption{
\label{fig:rnlsm2}
Time spent to compute the ionic gradients of the 
\protect{$\beta$}-projectors. 
Grey and blue lines, respectively, represent the routines and combined timing of 
the old implementation in CPMD 4.3.
Black and green lines, respectively, are the revised routines and combined timing 
of the new implementation. 
For {\tt rotate\_fnl} only half of the routine time is added to the combined time,
the other half contibutes to the gradients of the electronic states.
}

\end{figure}

The resulting timings for the ionic gradients due to the $\beta$-projectors
are shown in Figure~\ref{fig:rnlsm2}. As indicated above, 
the time spent in the new {\tt rotate\_fnl} and the revised {\tt rnlfl}
is almost negligible and both routines profit from {\tt cp\_groups} parallelization,
which is active at large core counts. 
The overall time now is dominated 
by {\tt rnlsm2}, which scales  perfectly up to very large core counts,
since no communication is involved. 
Similar to {\tt rottr\_fnl} in Fig.~\ref{fig:rnlsm1} the node local
routines {\tt rnlfl} and {\tt rotate\_fnl} contribute only very little to the overall timing.
Only in the scaling limit of our benchmarks going
beyond 3000 cores they
have a noticeable impact on the combined timings.

\subsection{Electronic Gradients of $\beta$-Projectors}
\label{ssec:nlforce}

Electronic gradients are required for every wave-function update,
either in a Car-Parrinello MD step or a Born-Oppenheimer MD  SCF cycle.
In CPMD the subroutine {\tt nlforce} calculates the non-local contributions to the
electronic gradient due to the $\beta$-projectors.

The implementation requires the same multiplication of $H$ to $F^\mathrm{NL}$ 
as in the {\tt rnlfl} routine. As indicated in  Sec.~\ref{ssec:Ions}, 
we use the result of  this multiplication from the  {\tt rotate\_fnl} routine 
and consider half its timing for the cost of the new implementation.
The remaining computational effort in {\tt nlforce} is a second matrix-matrix multiplication.
In the old implementation, this multiplication was called for every
$\beta$-projector of every atomic species, as in {\tt rnlsm1} and {\tt
  rnlsm2}. The new implementation uses a single compact matrix-matrix
multiplication combining {\em all} $\beta$-projectors, and, thus, makes best
use of thread and SIMD parallelization.

The matrix-matrix multiplication of the revised algorithm requires no communication, except
when the newly added  {\tt cp\_groups} parallelization is used.
Here, the matrix-matrix multiplication is distributed among the {\tt
  cp\_groups} and data is synchronized by an {\tt MPI\_allgather}. The cost
for the latter is reduced by an overlap of computation and communication,
similar to the {\tt rnlsm1} described in Figure~\ref{fig:rnlsm-flow}.

\begin{figure}
\includegraphics{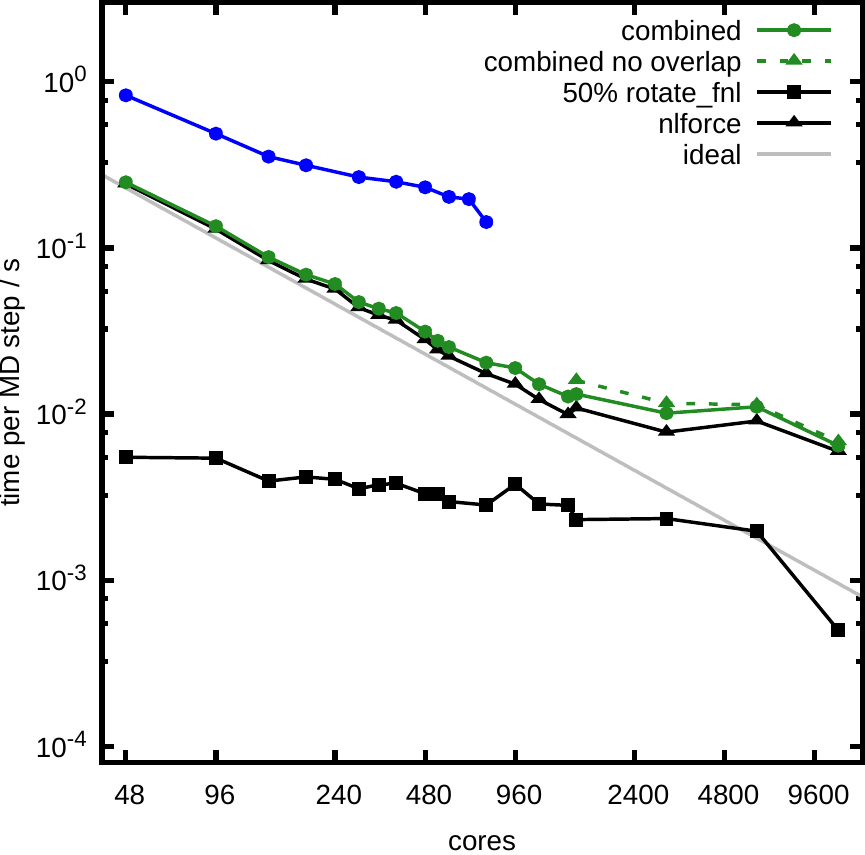}
\caption{
  The blue curve shows the timing for {\tt rnlfl} in the old implementation. In the new
  implementation reuses the quantity calculated in {\tt rotate\_fnl}. 
  The combined time of the new algorithm is given by the green solid line. 
  The green dashed line shows the
  combined time without overlapping communication and computation.
  The latter is only needed communication of the {\tt cp\_groups}
  parallelization. 
\label{fig:nlforce}
}
\end{figure}

The resulting speed up is demonstrated in Figure~\ref{fig:nlforce}. For a single node 
the new implementation is almost three times faster. 
It scales perfectly up to 1536 cores.
At this point {\tt cp\_groups} parallelization is activated and the effect of 
overlapping communication and computation to hide the synchronization
between {\tt cp\_groups} is visible and keeps the algorithm scaling up to 2400 cores.
At very high core counts the communication between {\tt cp\_groups} dominates
and  limits the scalability. Also the cost of {\tt rotate\_fnl} now
contributes noticeable to the overall timings, similar as mentioned in
Sec.~\ref{ssec:Ions}.

\subsection{Further USPP specific Optimizations}
\label{ssec:NLO}
Working through the list of time consuming routines, we identified four
additional routines of the USPP code path which consumed sizeable part of the
total compute time in the original CPMD version.
We have applied the same optimization paradigms as for
the routines above, namely targeting BLAS calls with large matrices,
complementing missing OpenMP directives and adding additional {\tt cp\_groups}
parallelization.

\begin{figure}
\includegraphics{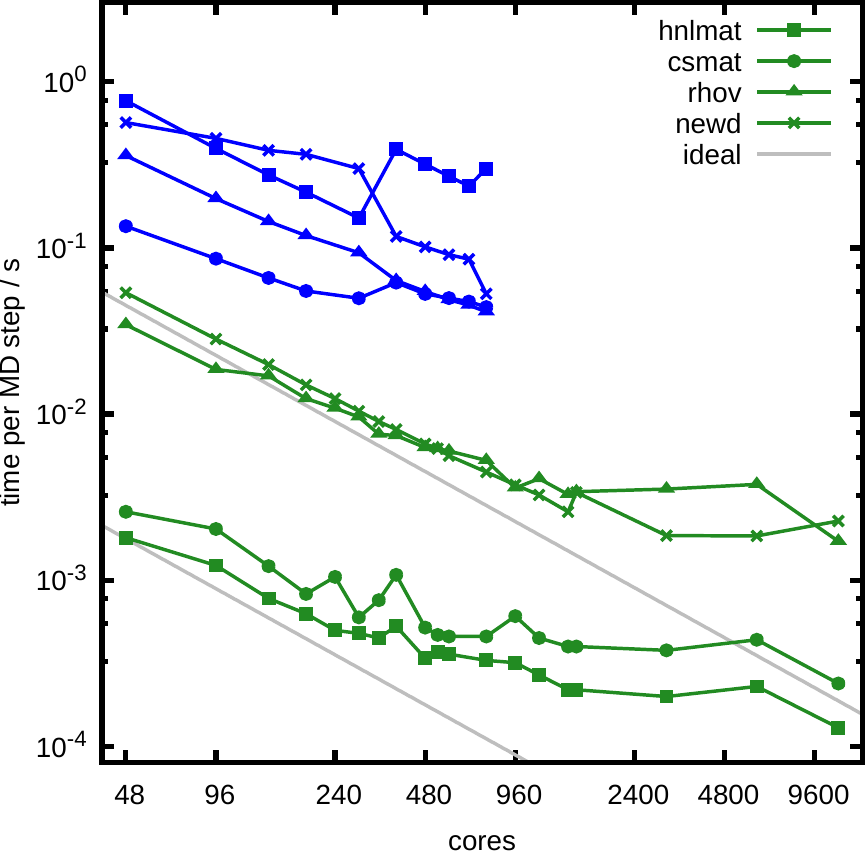}
\caption{ \label{fig:nlroutines}
Reduction of the time per MD step for various routines (symbols) comparing
CPMD 4.3 (blue lines) to the revised code version (green lines). }
\end{figure}

Figure \ref{fig:nlroutines} show the improvements of these four routines 
(green lines) with respect to CPMD 4.3 (blue lines). 
The routine {\tt newd} (cross symbols) computes the integral part of 
\eqref{eq:newd} and its ionic gradients, if required. Here, we now store 
the components of $Q^I$ instead of computing them on the fly in every time step.
Together with  improved matrix-matrix multiplications, this gives a speed up 
of about ten and the added {\tt cp\_groups} parallelization 
greatly improves the scalability.
Saving the $Q^I$ additionally speeds up the computation of 
the augmentation-charge density,
which adds to the square of the wave function in the computation of 
the total charge density in \eqref{eq:density}. The corresponding routine {\tt rhov}
also becomes significantly faster and 
scales similar as {\tt newd}.

The largest speed-up of almost three orders of magnitude was achieved for the
routine {\tt hnlmat} (square symbols), which computes the nonlocal part of the
Hamiltonian matrix $H$ \eqref{eq:hmat} stemming from the  sum expression in
\eqref{eq:hamiltonian}.
In the original algorithm the respective matrix elements have been updated in
the innermost part of a nested loop construct. Originally, the routine was MPI
parallelized by distributing the matrix elements of state pairs $i\le j$
across the MPI tasks.
In the revised implementation, we now distribute contributions of the
different atoms to the MPI tasks and compute the  partial contributions of an
atom to all state pairs within an MPI tasks. Thereby, we could replace a
complicated nested loop construct by a single and compute efficient {\tt
  DGEMM} call.
Now this routine is essentially negligible with respect to the total timing
($<0.1\,\%$).
Similarly, we have rewritten the {\tt csmat} routine (dot symbols), which
computes the overlap matrix $O$ \eqref{eq:csmat}. The resulting timings are
now comparable to the {\tt hnlmat} and equally do not play a role in the overall
timing any more.

\subsection{Batched Multi 3-$d$ FFT}
\label{ssec:fft}

One of the main computational bottlenecks in CPMD are the 3-$d$ FFTs  needed
to transform the electronic states (1024 for our 256 water benchmark) from
momentum space to real space.
The states in real space are needed twice per MD step, once for the calculation
of the electronic density \eqref{eq:density} and once for the interaction with 
the effective potential $V^\mathrm{eff}$ \eqref{eq:hamiltonian}. Subsequently, 
the product of the wave function with $V^\mathrm{eff}$ is transformed from real 
to momentum space to calculate the gradients on the electronic states. 
Since the coefficients of each state are distributed over the MPI tasks, 
an {\em all-to-all} communication is required for every single transform.  

In the continuous development of CPMD (and many other PW AIMD codes)
huge effort has been spent on the improvement of the 3-$d$ FFT algorithm. 
In CPMD the current implementation of  3-$d$ FFT treats only 
non-zero values explicitly in the FFT.\cite{Goedecker2003} As  back ends 
for the actual 1-$d$ FFT performance libraries such as FFTW or MKL 
can be used.\cite{FFTW05,MKL2019}

For the backward FFT of each of  the $N$ electronic states to real space, 
one needs first a FFT in $z$ direction, an {\tt MPI\_Alltoall} communication step, 
and then the FFTs in $y$ and $x$ direction. The execution of the FFT for 
the three stencil directions is the computationally demanding part, 
whereas the {\em all-to-all} communication is the major bottleneck for scalability.
Here, large MPI task counts lead to small message sizes which largely reduce 
the effective bandwidth of the communication.

A straight forward strategy for optimization is to have fewer communication steps
with larger  message sizes. For the multiple  3-$d$ FFTs we achieve this by packing
the communication from multiple electronic states, thus, implementing a
batched  multi 3-$d$ FFT algorithm.
Accordingly, we have rewritten all 3-$d$ FFT routines such that
each of the individual steps is executed for $b$ states and uses a buffer 
of $b$ times the size of the original one, where $b$ is the batch size.
Thereby, the message size of the {\em all-to-all} communication is $b$-fold
increased and the number of communication calls is $b$-fold reduced. 
Furthermore, also $b$-times as many stencils are transformed in a single call 
to the FFT library. This is  particularly beneficial in a hybrid MPI/OpenMP
parallelization scheme, where a threaded FFT is used and the individual FFTs 
are distributed among many cores.

\begin{figure}
\center{\includegraphics[width=5cm]{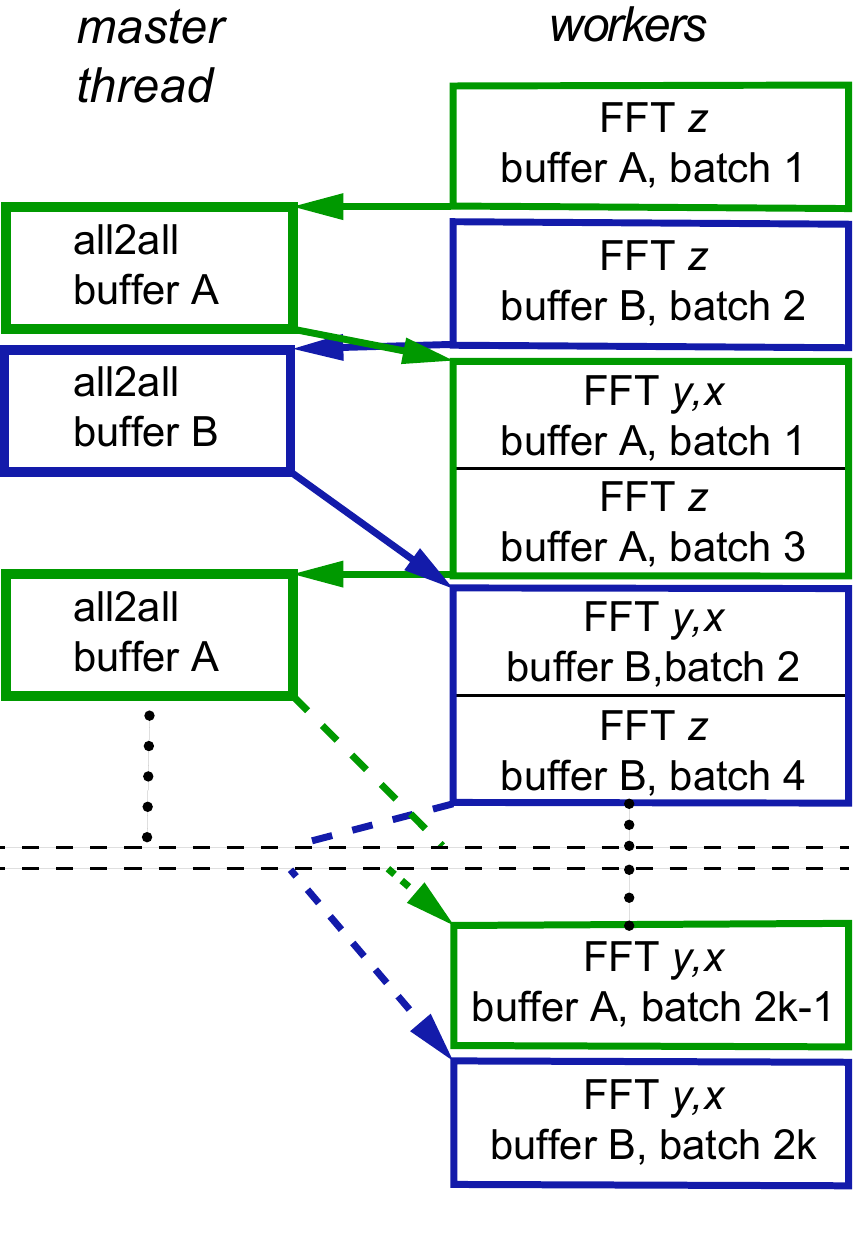}}
\caption{
\label{fig:fftflow}
 Program flow of the batched 3-$d$ FFT 
 with overlapping computation 
 and communication applied to  the electronic states from momentum space to real space.
 Batches with odd index use buffer $A$, 
 buffers with even index use buffer $B$.}
\end{figure}

Similar to the matrix-matrix multiplications in Sec.~\ref{ssec:ovl} we have, 
furthermore, extended the batched 3-$d$ FFT algorithm to overlap
computation and communication, see Figure~\ref{fig:fftflow}. 
Here, we use two buffers $A$ and $B$ that are  alternated for and 
even indexed batches.

After finishing the $z$ FFT of the first batch in buffer $A$ by the worker threads, 
the local master thread starts its  {\em  all-to-all} communication of buffer $A$.

Concurrent to this communication, the workers compute 
the $z$ FFT of batch 2 in  buffer $B$. 
After the master finishes the {\em all-to-all}
communication  of buffer $A$ control is handed over to the worker threads, 
which compute  the $y$ and $x$ FFTs of batch 1. Then the result is stored 
and buffer $A$ is refilled by the $z$ FFT of batch 3.  
Both master and worker threads alternate working on buffers $A$ and $B$.
As long as communication of a batch is shorter than 
the computation of its three fold FFT, the worker threads are constantly busy.

Choosing a sufficiently large batch size $b$  to saturate the bandwidth depends on
the simulation system and the machine it runs on. Here, we have implemented 
an auto-tuning algorithm that scans through the possible values of $b$ and
determines the fastest setting.
For our benchmark system we observe batch sizes from $b=1$ to $b=43$ with the
expected general trend that  for more  MPI tasks in each of the  {\tt
  cp\_groups} larger $b$ are preferred.

Finally, we omit the second call of the FFT to real space by storing the transformed
wave functions after the first call. This functionality has already been implemented
in CPMD but was inactive in the USPP code branch. For our benchmark system, 
the total memory required to store the wave functions is real space is 13.5\,GB.
Storing and retrieving this data will be  generally faster than the FFT it replaces  
but still is not negligible in the overall timing due to its size.

\begin{figure}
\includegraphics{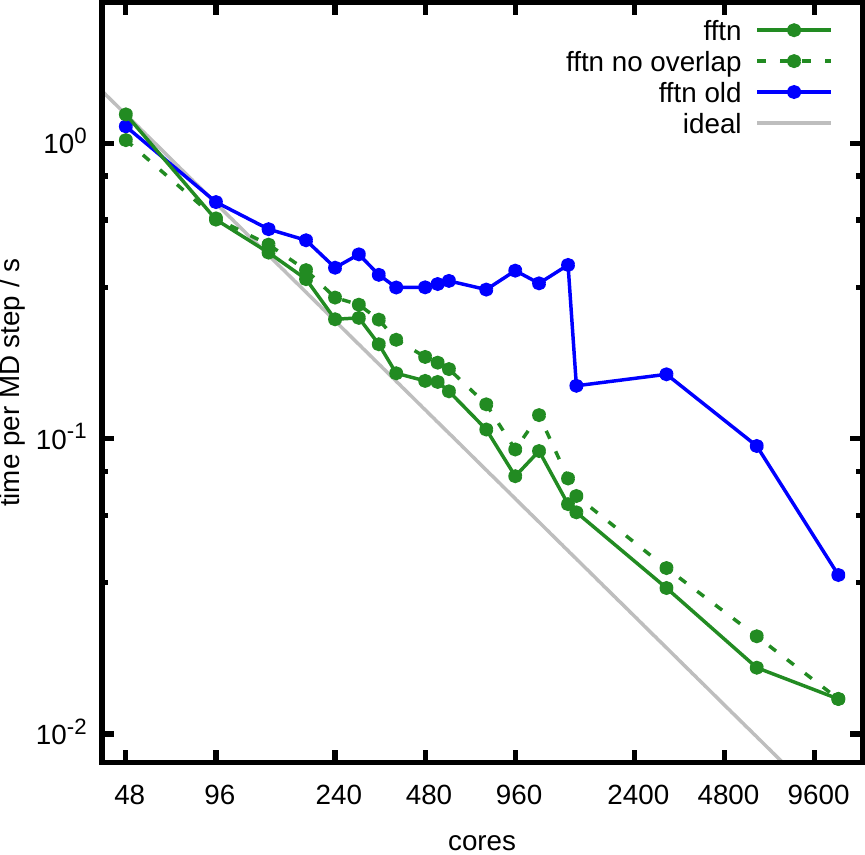}
\caption{
\label{fig:fft}
Total time spent in FFT with increasing core counts. Blue shows the old
implementation, green the new batched algorithm and green dashed the new
algorithm without overlapping communication and computation.}
\end{figure}

Since the original USPP implementation was not OpenMP parallel, we decided to
demonstrate the capabilities of the original FFT implementation using
our new implementation together with the old FFT implementation and compare
it to our new batched FFT routines in Fig.~\ref{fig:fft}.

The old implementation shows a good scalability up to 384 cores, where the
performance reaches a plateau. At 1536 cores
the effect of {\tt cp\_groups} parallelism is clearly visible, and enables
further scaling of the algorithm. Again we have to note that the
synchronization triggered by {\tt cp\_groups} parallelization is non-negligible as
already mentioned in Section \ref{ssec:ovl}.

Compared to the original implementation our batched FFT routines show better
scalability, if overlapping communication and computation is enabled,
the routine scales well up to 960 cores.
At this point the batched 3-$d$ FFT is  roughly a factor of 5 faster than the old
implementation.
At 1536 cores we can also observe a
speedup due to the {\tt cp\_groups}, however, the impact is much less than in the
original implementation. 
This execellent scaling without {\tt cp\_groups} parallelization
is needed up to  1536 cores avoids additional MPI
communication due to data synchronization
and has a large impact on the overall scalability of the code.
For larger node counts the new algorithm continues to scale up to 9600
cores and retains a sizable speedup compared to the old code. 

The effect of overlapping communication and computation 
is visible comparing the solid and dashed green lines.
It is in the
expected range, since one  thread exclusively handles the
communication. Thus the performance of the computational part is
reduced by a factor of $1/n$, where $n$ is the number of OpenMP threads. Since
in our benchmark system we found 12 OpenMP threads to be most efficient for
almost all core counts, the unavoidable performance loss is 8.3~\%. Yet we see
a overall performance improvement in the range of 20 - 30~\% indicating the
effectiveness of our implementation.

\subsection{USPP  Eigenvalue Problem}
\label{ssec:eigval}
Going to large node counts, one major scaling problem turned out to be 
solving of the symmetric
$N \times N$ eigenvalue problem in each MD
step to restore the orthogonality of the $\{|\psi_i\rangle\}$, see introduction.

Because the orthogonality is not constraint a unitary transformation matrix
needs to be constructed from the eigenvectors of the overlap\cite{Hutter95}

\begin{equation}
\label{eq:overlappsi}
\tilde O_{ij} = \langle \psi_i| \psi_j\rangle.
\end{equation}

For small systems, the associated computational effort is negligible
but for larger systems it is notable, particularly, as it  significantly hampers 
scaling at large core counts.

In the original code the LAPACK {\tt dspev} solver was executed on the
 MPI root rank and the eigenvectors were broadcasted to all MPI tasks.

We improved the scaling and the performance by replacing the packed LAPACK
routine {\tt dspev} with the unpacked version using 
the divide-and-conquer algorithm implemented in
the LAPACK routine {\tt dsyed}. 

This by itself lead to a significant speed up but 
the problem still constitutes a non-scaling part of the overall calculation. 

Employing the parallel Lanzcos method, which is already implemented in
CPMD \cite{BEKAS2008441},  failed to outperform the LAPACK routine, so we abandoned it. 

For systems of $N\gtrsim 2000$  the problem size is large enough 
that one can consider to use MPI-OpenMP parallel
eigenvalue solver, like EigenEXA \cite{EigenEXA}, ScaLAPACK \cite{ScaLAPACK99}, or ELPA.\cite{Marek2014}
We currently only have implemented an interface to the 
ELPA eigenvalue solver,  which is called on a subset of all available MPI
tasks. The achievable speed up, however, is of course strongly dependent on
the parallel setup and on the machine used.

Apart from speeding up the eigenvalue problem, the original implementation of
the USPP Car-Parrinello MD scheme applied this transformation only 
after a user defined threshold of the overlap (\ref{eq:overlappsi})
is reached \cite{Hutter95}.

We reimplemented the missing equations and also benchmarked the performance
impact of a threshold of $10^{-4}$ which is tighter than the recommended
threshold of $10^{-3}$ \cite{Hutter95}. In our benchmark calculations the
eigenvalue problem has to be solved only every 5th MD step. The corresponding
results are presented in Fig.~\ref{fig:weak} in Section \ref{ssec:performance}.

\section{Results and Discussion}
\label{sec:results}
All changes to the code introduced above resulted in an improved performance
and scalability. This performance is evaluated by the achievable trajectory
length measured in ps per day of walltime, which allows an easy estimate of  turnaround
times for a given simulation scenario. As a rule of thumb, a performance of
1\,ps/day is the minimum needed to achieve results on the order of a few
weeks. Below this limit, one will only conduct simulations only under
exceptional circumstances. On the other hand, performances largely exceeding
10\,ps/day will open the possibility to conduct many computer experiments on a
given system, due to the short turn-around times. We also provide the
performance in s/MD step. This quantity can be compared to a wave-function
update in an SCF cycle. Such a wave-function update is roughly 20~\% faster
than a Car-Parrinello MD step, since it does not involve calculations of ionic forces. 
A performance of 1\,ps/day equals 12.5\,s/MD step or roughly 10\,s/SCF
cycle. For very large systems exceeding 512 water molecules, this is a better
quantity as here most DFT calculations are conducted to optimize geometries or
even just to analyze the band structure of a given system. 

\subsection{Performance and Scalability  of the 256 Water Benchmark System}
\label{ssec:performance}
First, we turn to the benchmark system of 256 water molecules, for which we
have already discussed the improvements of the subroutines in Sec.~\ref{sec:optimization}. 
Table \ref{tab:results} shows the timings and setup (OpenMP threads, {\tt 
cp\_groups}) with the best performance for the revised CPMD code.
With the improved threading support and the {\tt cp\_groups} parallelization we 
achieve a  performance increases for up to 240 nodes (11.5\,k cores).
For most setups 12 OpenMP threads per MPI task yield the best performance, which 
corresponds to 4 MPI tasks per node. Only for small node counts
and the limiting 240 nodes a smaller and larger number of threads, respectively, 
is preferential. The {\tt cp\_groups} parallelization
is profitable only for more than 30 nodes and is required to use very large core
counts.
\begin{table}
  \label{tab:results}
  \caption{For a given number of nodes the table lists the OpenMP threads and the number of {\tt cp\_groups} $g$ 
           yielding the shortest time per MD step and the corresponding best performance.}
 \begin{tabular}{rrrcSS}
   nodes       &   cores       &    threads     & {$g$}  &   {time}      &   {perf.}   \\
               &               &                &        &     {s}                 & {ps / day} \\
  \hline
     1         &    48         &     4      &     1             &3.61             &  3.47        \\
     2         &    96         &     4      &     1             &1.84             &  6.79        \\
     3         &   144         &     6      &     1             &1.31             &  9.57        \\
     4         &   192         &     8      &     1             &1.03             &  12.1        \\
     5         &   240         &     6      &     1             &0.840            &  14.9        \\
     6         &   288         &    12      &     1             &0.738            &  17.0        \\
     7         &   336         &     8      &     1             &0.647            &  19.4        \\
     8         &   384         &     6      &     1             &0.569            &  22.0        \\
    10         &   480         &    12      &     1             &0.500            &  25.1        \\
    11         &   528         &    12      &     1             &0.472            &  26.6        \\
    12         &   576         &    12      &     1             &0.439            &  28.5        \\
    16         &   768         &    12      &     1             &0.345            &  36.4        \\
    20         &   960         &     8      &     1             &0.316            &  39.6        \\
    24         &  1152         &    12      &     1             &0.300            &  41.8        \\
    30         &  1440         &    12      &     1             &0.250            &  50.1        \\
    32         &  1536         &    12      &     2             &0.228            &  54.9        \\
    64         &  3072         &    12      &     2             &0.167            &  75.2        \\
   128         &  6144         &    12      &     4             &0.140            &  89.6        \\
   240         & 11520         &    24      &     8             &0.130            &  96.5    
   \end{tabular}
\end{table}

\begin{figure}
\includegraphics{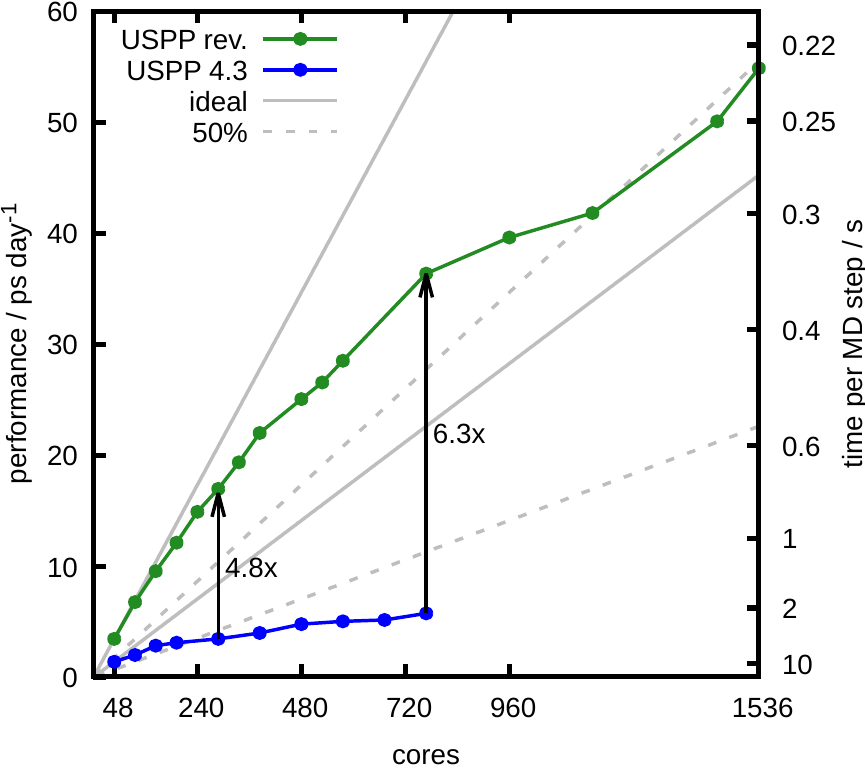}
\caption{
\label{fig:psperday}
Performance improvement for the 256 water benchmark system with USPP. Blue
shows the old and green the new implementation. 50~\% efficiency is shown in
grey dashed. Please note the linear scale of both axes.}
\end{figure}

Figure~\ref{fig:psperday} shows the scaling data of the revised code given in Tab.~\ref{tab:results} and compares it to the performance of
the original code on a linear scale.

%
 
For a single node (48 cores) the revised implementation outperforms the  old code 
by a factor of 2.5. Up to six nodes, the old code is faster in a pure MPI set up and
profits from hybrid MPI+OpenMP parallelization only for larger node counts. However, 
the parallel efficiency drops significantly below 50\,\%, which is often considered a threshold
to use the HPC facilities at computing centers.
 
In contrast, the hybrid  MPI+OpenMP parallelization of the new implementation is faster
for all node counts. On a single node it uses 4 OpenMP threads per MPI task 
and 12 OpenMP threads per task at 20 nodes (960 cores).
Its performance at 6 nodes is 4.8 times faster and
 16 nodes it is more than 6 times faster.
The parallel efficiency stays well above a 50\,\% up to 20 nodes. 

At 30 nodes (1440 cores) the efficiency
drops slightly below 50\,\%, however, at 32 nodes (1536 cores) it reclaims
the 50\,\% efficiency barrier by using {\tt cp\_groups} parallelism. 
The achieved performance is about 55\,ps/day or less than 0.23\,s per MD
step. Compared to the old code, this is more than 9 times faster by using just
twice the resources.

\begin{figure}
\includegraphics{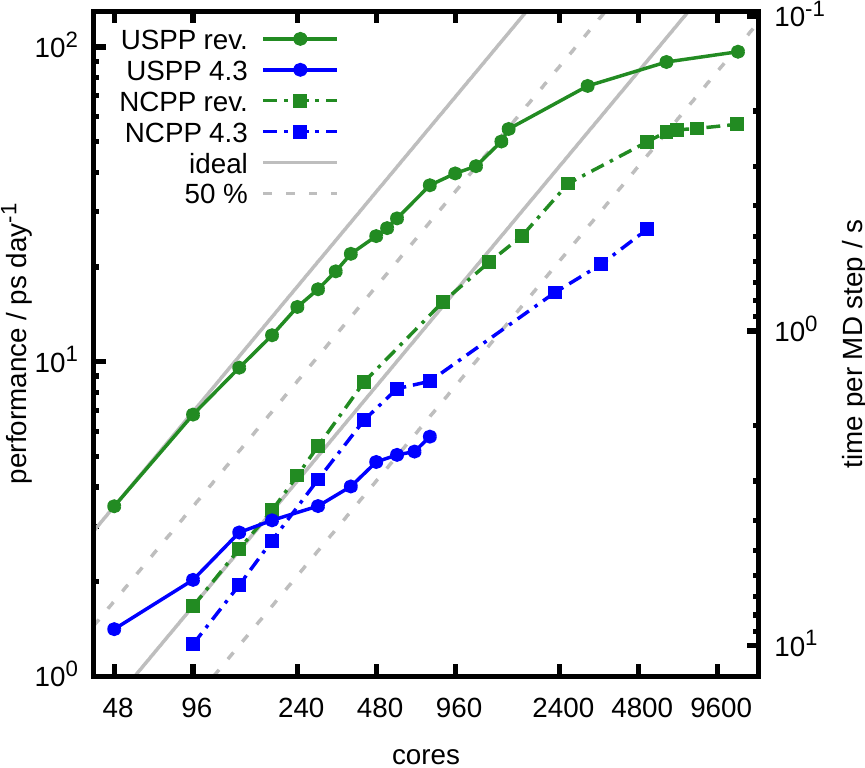}
\caption{
\label{fig:psperdaync}
Performance of USPP (solid) and NCPP (dashed) simulations. Blue lines show the
old and green the new implementation.}
\end{figure}

Despite our focus on  the USPP code path, also  NCPP calculations profit
 from our optimizations, particularly due to the batched FFT, 
 see Section \ref{ssec:fft}.
The log-scaled Figure~\ref{fig:psperdaync} compares the performance of the original 
and the revised code for USPP and NCPP, respectively, on the 256 water benchmark system.
At low node counts up to 12 the performance improvement of the NCPP code path is 
30\,\% on average. For larger node counts the new batched FFT algorithm 
plays its strength and retains perfect scaling much longer than
 the old implementation.
Here, the performance improvement is more than 50\,\% and  the parallel efficiency
is larger than 50\,\% even at 108 nodes  (5184 cores).

The old USPP implementation was only up to 4 nodes faster that NCPP. Note that for 
NCPP on a single node the memory was insufficient to store the  
wave-function FFTs, which are six times larger that for USPP 
due to the larger PW  cutoff. 
Therefore, we have omitted this point for NCPP and use two nodes for its reference. 
For the minimal configuration USPP is now  4.5 times more efficient than NCPP in the new implementation, 
which covers about 80\,\% of the  maximum possible speed up 
of 5.7 due to the ratio of the number of PW.
Thus, the USPP overhead related
to the larger number of $\beta$-projectors and additional terms that have to be computed 
now seems to be in a well acceptable limit.
The scaling of USPP is not as close to ideal as NCPP and  drops just below
50\,\% parallel efficiency after 32 nodes (1536 cores). Nevertheless, it remains
faster than than NCPP, outperforming it by 70\,\% above 5000 cores. 

\subsection{System Size Dependence}
Finally, we consider the scalability and performance  depending on  a broad
range of system sizes. The smallest system of 32 molecules was used as a
minimal system to study liquid water and aqueous solutions in the early days
of AIMD \cite{Tuckerman95,Marx99}.
The largest system of 2048 water molecules covers  8196 electronic states.
For similarly sized systems  structure optimizations at DFT level are of interest today.

\begin{figure}
\includegraphics{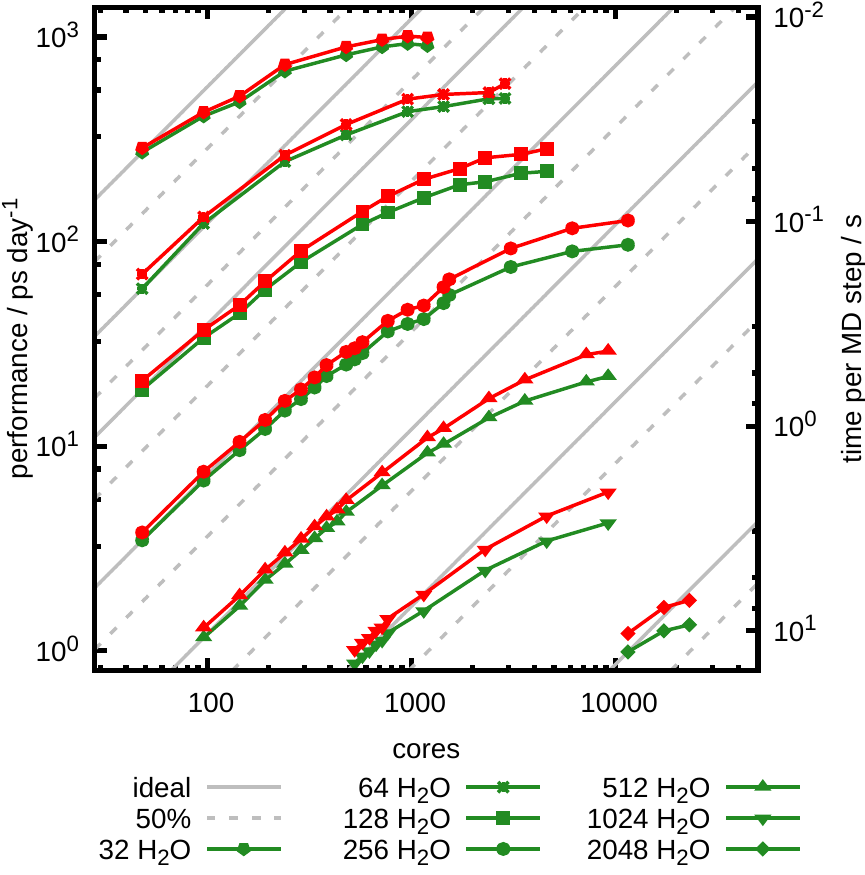}
\caption{
\label{fig:weak}
Performance and scaling of the revised CPMD USPP code for water systems of
increasing size (green curves). Minimal core set ups for each system have been chosen due to
memory restrictions. Red lines show the impact of additionally reimplementing the threshold
to trigger the eigenvalue problem.}
\end{figure}

All scaling curves are collected in Figure~\ref{fig:weak}. For the smallest
system of 32 molecules we reach a performance of more than 1\,ns/day with 20
nodes, a performance scale previously only known from force-field based MD
simulations.  Of course, the parallel efficiency at large core counts is
rather poor due to the small size of the system.

For 64 molecules one readily reaches more than 264\,ps/day with 5 nodes (240
cores)  at a good parallel efficiency. The next system, consisting of 128
molecules we reach a performance of 121\,ps/day with just 12 nodes before
dropping below 50\,\% parallel efficiency. In the scaling limit we can reach a
performance exceeding 590\,ps/day and 284\,ps/day respectively.
A system twice the size of our benchmark system at 512 molecules
requires 1440 cores to reach more
than 10\,ps/day with  while the maximum achievable performance is
greater than 20\,ps/day.

Very large systems test the limits of the implementation. At 1024
molecules we find a good scaling up to 2304 cores reaching about 2,5\,ps/day
and more than 7\,ps/day in the scaling limit. This shows that our code
revisions elevate this system size to a performance sufficient for production
runs. Starting with simulations of this size, we also see additional
subroutines hampering scaling that do not play a role for smaller systems and 
which leave room for further improvements.
One
example is the calculation of the inverse root $T=O^{-1/2}$ of
\eqref{eq:csmat}. $T$ is calculated as the inverse of a Cholesky factor by
subsequent calls to the Lapack routines {\tt dpotrf} and {\tt dtrtri} on the
MPI root rank. A possible enhancement would be the adaption of the blocked Gram-Schmidt
orthogonalization method implemented by Bekas for NCPP to the USPP formalism. \cite{BEKAS20101057}

For the largest benchmark system of 2048 molecules we found only 
three stable simulation setups, because the memory requirements were exceeding the
available node memory of 96\,GB. Yet we were able to reach a performance of more than
1.7\,ps/day, making it accessible for production runs.

Particularly these large systems are hampered by solving the eigenvalue problem
described  in Sec.~\protect{\ref{ssec:eigval}}.
The red curves show the advantage of restoring the original Hutter
implementation where the respective function is called on average only every 5th MD step
due to the active treshold.
Since the eigenvalue problem too small 
to 
be as efficiently parallelized as the rest of the  code
its impact is most pronounced in the scaling limit,
as Amdahl's law predicts.\cite{Amdahl}
For the larger systems, the speedup is on the order of up to 30\,\%.

Note that DFT does not scale linearly with the system size $N$ but rather
with $N^\alpha, \alpha>1$, From respective  minimal configurations 
presented in Fig.~\ref{fig:weak} we calculated an
effective  single node performance for each system size.
We find that the USPP code path scales with $\sim
N^{2.6}$.

Compared to our intermediate results acquired on SuperMUC-Phase2\cite{posterSC18} 
we were able to adopt our routines to the new SuperMUC-NG and
obtain now at least twice the performance for small systems and
close to ten times higher performance with the largest benchmark system of 2048
water molecules. The scalability improvement was mainly achieved by the now
communication free {\tt rnlsm2} routine described in Section \ref{ssec:Ions} whereas
the improved core performance was just obtained by using highly optimized BLAS
routines all over the code at did not require any further tuning on our side.

\section{Summary and Outlook}
Optimizing AIMD  for HPC systems is a challenging task despite its high
computational intensity. One  obstacle is  the large amount of data needed to describe 
the electronic states, which needs to be distributed among the nodes. Still,
the electronic states are tightly coupled requiring  {\em all-to-all} data exchange between the nodes,
which is a large disadvantage compared to scalable nearest neighbour communication
used by many grid based codes.
Also instead of a single monolithic kernel, AIMD has many inter-dependent compute tasks.
Due to the data distribution each task requires a different parallelization  strategy which makes
the whole optimization process particularly tedious.

We have greatly enhanced the performance and scalability of the USPP
AIMD implementation in CPMD. All rate limiting routines have been revised and
improved and communication patterns have been reworked. For a number of
routines we have shown that overlapping computation and
 communication is indeed profitable when reaching the scaling limit.
In the revised code the distributed matrix-matrix multiplications in
the {\tt rnlsm} routine and the 3-$d$ FFTs 
still require the largest computational effort and determine the limits of scalability.
The algorithmic changes we have introduced have  pushed these limits farther out
and allow more cores to  work on a given problem. 

With the code improvements we have not only increased the reasonably
accessible system sizes for CPMD simulations by at least one order of
magnitude along with much shorter turn-around times but also largely stretched
the accessible time range, bringing multi nanosecond CPMD simulations down to
just a few days of simulation for systems up to 256 water molecules.

Further development will be on the methods side, implementing Hartree Fock
exchange \cite{hfx} together with wave-function localization
methods\cite{SCDM,MLWF}. Due to our high usage of performance libraries, we
are also confident that porting our routines to heterogenous HPC cluster with
accelerators will be straight forward.

\section*{Acknowledgments}
This work has been funded by state of Bavaria (KONWIHR software
initiative). The authors gratefully acknowledge the compute resources and
support provided by the Erlangen Regional Computing Center (RRZE). Technical
support for optimization has been granted by the PRACE High Level Support team
(HLST) at the Leibniz Supercomputing Centre (LRZ). Computer time for
benchmarking was generously provided by the LRZ during the reliability-testing
phase of SuperMUC-NG (project {\tt pn69qo}).


\end{document}